\documentclass[12pt,aps,prd,preprint,tightenlines,
   showpacs,nofootinbib]{revtex4}
\newcommand{\PRE}[1]{{#1}} 

\usepackage{bm} \usepackage{epsfig}
\usepackage{slashed}
\usepackage{multirow}

\newcommand{\gev}{\text{GeV}}
\newcommand{\tev}{\text{TeV}}
\newcommand{\fb}{\text{fb}}

\newcommand{\eqref}[1]{Eq.~(\ref{#1})}
\newcommand{\eqsref}[2]{Eqs.~(\ref{#1}) and (\ref{#2})}
\newcommand{\secref}[1]{Sec.~\ref{sec:#1}}

\newcommand{\figref}[1]{Fig.~\ref{fig:#1}}

\newcommand{\tableref}[1]{Table~\ref{table:#1}}

\newcommand{\bea}{\begin{eqnarray}}
\newcommand{\eea}{\end{eqnarray}}

\newcommand{\LambdanotU}{\ensuremath{\Lambda_{\slashed{\mathcal{U}}}}}
\newcommand{\LambdaU}{\ensuremath{\Lambda_{\mathcal{U}}}}
\newcommand{\duv}{d_{\text{UV}}}
\newcommand{\Ouv}{O_{\text{UV}}}

\newcommand{\nolabel}{{}}

\begin{document}

\preprint{UCI-TR-2007-25}

\title{
\PRE{\vspace*{1.5in}} Unparticles: Scales and High Energy Probes
\PRE{\vspace*{0.3in}} }

\author{Myron Bander}
\affiliation{Department of Physics and Astronomy, University of
California, Irvine, CA 92697, USA \PRE{\vspace*{.5in}} }

\author{Jonathan L.~Feng}
\affiliation{Department of Physics and Astronomy, University of
California, Irvine, CA 92697, USA \PRE{\vspace*{.5in}} }

\author{Arvind Rajaraman}
\affiliation{Department of Physics and Astronomy, University of
California, Irvine, CA 92697, USA \PRE{\vspace*{.5in}} }

\author{Yuri Shirman%
\PRE{\vspace*{.2in}} } \affiliation{Department of Physics and
Astronomy, University of California, Irvine, CA 92697, USA
\PRE{\vspace*{.5in}} }

\date{June 2007}

\begin{abstract}
\PRE{\vspace*{.3in}} Unparticles from hidden conformal sectors provide
qualitatively new possibilities for physics beyond the standard model.
In the theoretical framework of minimal models, we clarify the
relation between energy scales entering various phenomenological
analyses.  We show that these relations always counteract the
effective field theory intuition that higher dimension operators are
more highly suppressed, and that the requirement of a significant
conformal window places strong constraints on possible unparticle
signals.  With these considerations in mind, we examine some of the
most robust and sensitive probes and explore novel effects of
unparticles on gauge coupling evolution and fermion production at high
energy colliders.  These constraints are presented both as bounds on
four-fermion interaction scales and as constraints on the fundamental
parameter space of minimal models.
\end{abstract}


\maketitle

\section{Introduction}

Among the many candidates for new physics are hidden sectors coupled
to the standard model through non-renormalizable interactions
\begin{equation}
\frac{\Ouv \, O_{\text{SM}}}{M^{m+n-4}} \ ,
\label{interactions}
\end{equation}
where $\Ouv$ and $O_{\text{SM}}$ are hidden sector and standard model
operators with mass dimensions $m$ and $n$, respectively.  $M$ is the
energy scale characterizing the new physics, which may range from the
weak scale to the Planck scale.  Hidden sectors that become either
weakly coupled or strongly coupled at low energies have several
interesting motivations and have been well-studied from various
viewpoints.

Recently a novel possibility was introduced in
Refs.~\cite{Georgi:2007ek,Georgi:2007si}, which suggested that the
hidden sector could be conformal at an energy scale $\LambdaU$.
Conformal hidden sectors have bizarre implications, including, for
example, kinematic distributions in the production and decay of
standard model particles that have no conventional particle
interpretation.  This possibility is therefore qualitatively different
from other candidates for new physics, and ``unparticles,'' the
degrees of freedom of the conformal sector, have recently been the
subject of several phenomenological
studies~\cite{Cheung:2007ue,Luo:2007bq,%
Chen:2007vv,Ding:2007bm,Liao:2007bx,Aliev:2007qw,Lu:2007mx,%
Li:2007by,Duraisamy:2007aw,Stephanov:2007ry,Fox:2007sy,%
Greiner:2007hr,Davoudiasl:2007jr,Choudhury:2007js,Chen:2007qr,%
Aliev:2007gr,Mathews:2007hr,Zhou:2007zq,Ding:2007zw,Chen:2007je,%
Liao:2007ic}.

Depending on the properties of the unparticle operators, the general
interactions of \eqref{interactions} may generate many specific
non-renormalizable interactions when $\Ouv$ flows at low energies to
an operator $O$ with mass dimension $d$.  As argued
in~\cite{Georgi:2007ek,Georgi:2007si}, these interactions, such as
\begin{equation}
\frac{O H^2}{\Lambda^{d-2}} \, , \
\frac{O^{\mu} \overline{f} \gamma_{\mu} f}{\Lambda^{d-1}} \, , \
\frac{O H\overline{f} f }{\Lambda^d} \, , \
\frac{O F_{\mu\nu} F^{\mu\nu}}{\Lambda^d} \ ,
\label{operators}
\end{equation}
have implications for a plethora of experiments, and lower bounds on
the scales $\Lambda$ have already been derived by considering a wide
range of topics, from anomalous magnetic moments to CP violation to
production rates at high energy colliders.

In this paper, we begin by investigating the theoretical framework of
unparticles from a general point of view.  We note that the scales
$\Lambda$ appearing in \eqref{operators} are not identical or even
generically comparable.  In fact, when expressed in terms of the
fundamental energy scales $M$ and $\LambdaU$, the scales $\Lambda$ of
\eqref{operators} are typically hierarchically separated, and this
hierarchy always counteracts the standard intuition from effective
field theory that operators suppressed by more powers of $\Lambda$ are
less important. Following the work of
Refs.~\cite{Georgi:2007ek,Georgi:2007si}, we explore minimal
unparticle models to provide simple frameworks for phenomenological
studies.  This approach clarifies certain issues.  For example, some
phenomenological observables become sensitive to arbitrarily high
scales $\Lambda$; we show that this sensitivity is artificial, and
there is no singularity when bounds are expressed in terms of the
fundamental parameters $M$ and $\LambdaU$.

Another essential point is that the conformal symmetry does not
generically hold to arbitrarily low energies~\cite{Fox:2007sy}. In
fact, the first operator of \eqref{operators} breaks conformal
invariance at a scale $\LambdanotU$~\cite{Fox:2007sy}.  Experimental
probes of the conformal hidden sector must probe energies in the
conformal window $\LambdanotU < E < \LambdaU$.  As we will see, this
criterion is very restrictive. In natural models, which we define more
precisely below, it implies that only experiments at energies near the
weak scale $v \simeq 246~\gev$ are viable probes of conformal hidden
sectors. Furthermore, requiring a reasonably wide conformal window
sets $\LambdaU \ll M$, which implies that all the couplings of the
unparticle sector to the standard model are extremely suppressed and
there are no accessible experimental signatures. Experimental
signatures in a significant conformal window are possible only if the
first coupling of \eqref{operators} is absent, for reasons we discuss,
or is fine-tuned to unnaturally small values.

Last, we examine in detail several leading constraints on unparticle
physics, considering both scalar unparticles and vector unparticles in
turn.  Given the considerations noted above, we focus on high energy
probes, which are most likely to be in the conformal window. In
particular, we consider bounds from $e^+e^-$ colliders in the energy
range $30~\gev$ to $200~\gev$, and derive bounds on the scales
$\Lambda$ in \eqref{interactions} and also their implications for the
fundamental parameters $M$ and $\LambdaU$.  We also find a novel
signature of scalar unparticles. These operators lead to a coupling
between the Higgs field and gauge bosons, which in turn leads to a
modification of the gauge couplings that currently provides a severe
constraint and is in the future potentially observable.

We close with a summary of our main results and outline directions for
further work.

\section{Theoretical Framework}

\subsection{Scales}
\label{sec:scales}

Following Refs.~\cite{Georgi:2007ek,Georgi:2007si}, we assume that in
the ultraviolet theory, a hidden sector operator $\Ouv$ with dimension
$\duv$ couples to standard model operators $O_n^i$ with dimension $n$
through the coupling
\begin{equation}
c_n^i \frac{\Ouv O_n^i}{M^{\duv + n -4}} \ .
\label{coupling}
\end{equation}
The hidden sector becomes conformal at energy $\LambdaU$, and the
operator $\Ouv$ flows to an operator $O$ with dimension $d$.  At low
energies, then, the couplings of \eqref{coupling} flow to
\begin{equation}
c_n^i \frac{O O_n^i \LambdaU^{\duv-d}}{M^{\duv+n-4}}
\equiv c_n^i \frac{O O_n^i}{\Lambda_n^{d+n-4}} \ .
\end{equation}
The scales $\Lambda_n$ determine the strengths of the couplings
between the unparticle operator and standard model operators of
dimension $n$.

We emphasize that operators of different dimensions couple with
different strengths. For example, standard model fermions couple to
vector unparticles through interactions
\begin{equation}
c_3^{f_i f_j} \frac{O^{\mu} \overline{f_i} \gamma_{\mu} f_j}
{\Lambda_3^{d-1}}
\end{equation}
suppressed by $\Lambda_3$.  (Note that these operators become almost
renormalizable for $d$ near 1.)  On the other hand, standard model
gauge bosons couple to scalar unparticles through interactions
\begin{equation}
c_4^{F^i} \frac{O F^i_{\mu\nu} F^{i\,\mu\nu}}{\Lambda_4^d}
\end{equation}
suppressed by $\Lambda_4$, where $i$ labels the gauge group.  Standard
model fermions may also couple to scalar unparticles through an
interaction derived from a 4-point coupling when the Higgs boson gets
a vacuum expectation value (vev) $\langle H \rangle = v \simeq
246~\gev$ after electroweak symmetry breaking:
\begin{equation}
c_4^{f_i f_j} \frac{O H\overline{f_i} f_j}{\Lambda_4^d} \to 
c_4^{f_i f_j} \frac{O v \overline{f_i} f_j}{\Lambda_4^d} \equiv
c_4^{f_i f_j} \frac{O \overline{f_i} f_j}{\Lambda'_3{}^{d-1}} \ ,
\end{equation}
where the last form defines another scale $\Lambda'_3$, which is
sometimes constrained in phenomenological studies.  Finally, the Higgs
couples through the operator
\begin{equation}
c_2 \Lambda_2^{2-d} O H^2 \ .
\label{Higgs}
\end{equation}
This coupling is special in many contexts, as it is the unique
super-renormalizable coupling to gauge singlet new
physics~\cite{Strassler:2006im,Patt:2006fw}.  In the present context,
it also plays an essential role, because once the Higgs develops a
vev, this operator breaks conformal symmetry in the hidden
sector~\cite{Fox:2007sy}.  This breaking occurs at the scale
\begin{equation}
\LambdanotU = \left(c_2 \Lambda_2^{2-d} v^2\right)^{\frac{1}{4-d}} \ .
\end{equation}

\subsection{Minimal Models}
\label{sec:models}

To unify the many scales and couplings discussed above, we may assume
a single unparticle operator $\Ouv$ coupling to the standard model.
This assumption of a single unparticle operator defines minimal
unparticle models, which are fully specified by the fundamental
parameters
\begin{equation}
\left\{S , \ M , \ \LambdaU , \ \duv , \ d , \ c_2 \
[ , \ c_{n \ge 3}^i ] \ \right\} \ ,
\end{equation}
where $S$ is the spin of the hidden sector operator $O$, and all other
parameters are defined in \secref{scales}.  The many parameters $c_{n
\ge 3}^i$ enter into interaction terms.  However, as we will see
below, very few enter any given process, and for the purposes of
setting bounds consistent with conventions in the
literature~\cite{Eichten:1983hw,Yao:2006px}, we may simply set $|c_{n
\ge 3}^i| = \sqrt{2\pi}/e$.

Neglecting the $c_{n \ge 3}^i$, the minimal model is defined by the
discrete parameter $S$, two fundamental scales $M$ and $\LambdaU$, and
three continuous parameters, $\duv$, $d$, and $c_2$.  These
fundamental parameters completely determine the couplings of
unparticles to the standard model, and provide a complete, yet simple,
framework for studying the phenomenology and cosmology of unparticles
and hidden conformal sectors.

It is convenient to define dimensionless ratios
\begin{equation}
r \equiv \frac{\LambdaU} {M} \leq 1 \qquad 
s \equiv \frac{\LambdaU}{v} \,.
\end{equation}
All remaining mass scales of the minimal model are, then, related to
the mass scale $\LambdaU$ through
\begin{eqnarray}
\Lambda_2 &=& r^{\frac{\duv-2}{2-d}} \LambdaU \\
\Lambda_3 &=& \left( \frac{1}{r} \right)
^{\frac{\duv-1}{d-1}} \LambdaU \label{Lambda3} \\
\Lambda_4 &=& \left( \frac{1}{r} \right)
^{\frac{\duv}{d}} \LambdaU \label{Lambda4} \\
\Lambda'_3 &=& \left( \frac{1}{r} \right)^{\frac{\duv}{d-1}}
s^{\frac{1}{d-1}} \LambdaU \label{Lambda3prime} \\
\LambdanotU &=& \left( \frac{c_2 r^{\duv-2}}{s^2}
\right)^{\frac{1}{4-d}} \LambdaU \ ,
\label{lambdanot}
\end{eqnarray}
where we have written all expressions with positive exponents,
assuming $1 < d < 2 < \duv$.

From these expressions, we see that $\Lambda_2 < M < \Lambda_4 <
\Lambda_3$.  This hierarchy (partially) offsets the standard intuition
of effective field theories that operators suppressed by more powers
of the characteristic energy scale are less promising to probe.  For
example, $\Lambda_4 < \Lambda_3$ implies that gauge couplings probed
by, for example, gluon-gluon collisions at hadron colliders, may yield
more promising signals than couplings to fermions.

Note that we have assumed $r \le 1$, that is, that the theory enters
the conformal regime below the scale where it is coupled to the
standard model. This is required for theoretical consistency. If
$\LambdaU>M$, then the theory is already conformal at the scale $M$,
and so the effective couplings $\Lambda_n$ should be replaced simply
by $M$. This is equivalent to taking $\LambdaU = M$ in all the
formulae, and so we may take $\LambdaU\leq M$ without loss of
generality.

\subsection{Naturalness}

As argued in Ref.~\cite{Fox:2007sy}, the breaking of conformal
invariance at energy scale $\LambdanotU$ means that unparticle physics
is relevant only for experiments that probe energies $\LambdanotU < E
< \LambdaU$.  From \eqref{lambdanot}, we see that creating a conformal
window spanning, say, one order of magnitude requires
\begin{equation}
\frac{c_2 r^{\duv-2}}{s^2} < \left(\frac{1}{10}\right)^{4-d} \ .
\label{criterion}
\end{equation}
This is a significant constraint, given $1<d<2$.

One way to satisfy \eqref{criterion} is to take large $s$, that is,
$\LambdaU \gg v$.  This raises the energy scale of all unparticle
interactions and rapidly decouples unparticles from accessible
experiments.  For unparticles to be accessible through weak-scale
experiments, one must take $\LambdaU \sim v$.  (Note that precision
experiments, for example, those probing flavor, CP, and baryon number
violation, probe scales far above the weak scale; however, these are
typically conducted at very low energies outside the conformal
window.)

\begin{figure}
\resizebox{3.65 in}{!}{\includegraphics{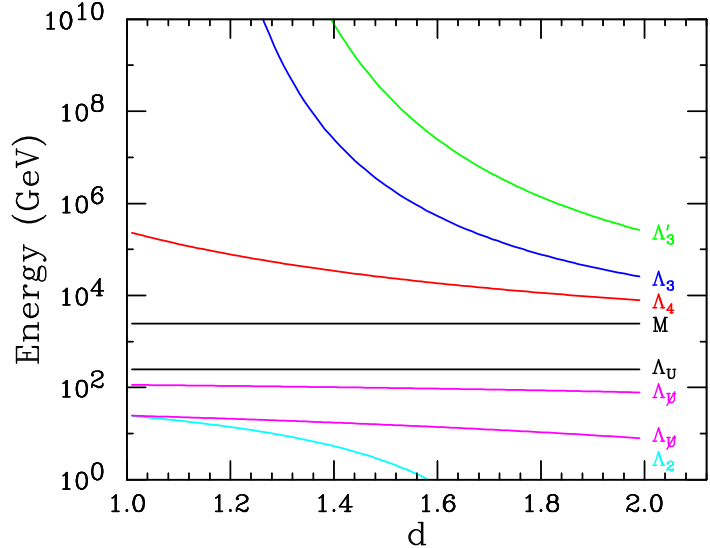}} 
\caption{Energy scales in the minimal unparticle model as functions of
$d$, assuming $\LambdaU = v \simeq 246~\gev$, $M=10v$, and $\duv =
3$. The two lines for $\LambdanotU$ are for $c_2=1$ (upper) and
$c_2=0.01$ (lower).
\label{fig:scales10}
}
\end{figure}

A second option is small $r$, that is, $M \gg \LambdaU$.  As evident
from Eqs.~(\ref{Lambda3}), (\ref{Lambda4}), and (\ref{Lambda3prime}),
this raises $\Lambda_3$, $\Lambda_4$, and $\Lambda'_3$ rapidly, again
making unparticle physics inaccessible.  This is illustrated in
\figref{scales10}, where we have plotted all relevant energy scales as
functions of $d$ for $\LambdaU = v$, $\duv =3$, and $M = 10 \LambdaU$.
Even for this relatively small hierarchy between $M$ and $\LambdaU$,
which creates only a slight conformal window, we find $\Lambda_4,
\Lambda_3, \Lambda'_3 \agt 10~\tev$, which is likely beyond the reach
of foreseeable experiments.

The third and final logical possibility is $c_2 \ll 1$.  Naturalness
suggests $c_2 \sim 1$. Furthermore, similar to quadratic divergences
in the Higgs boson mass, quantum corrections to $c_2$ have a power law
divergence, which scales as $\LambdaU^{2-d}$. Thus even if one sets
$c_2=0$ at tree level, it becomes of order $\lambda_t c_t/(16\pi^2)
\sim 0.01$ at loop-level.  Of course, if there are no scalar
unparticle operators $O$ with dimension $d<2$, then $c_2 = 0$.
Consideration of scalar operators with $d >2$ requires extending the
standard range $1<d<2$ through the singularities at $d=2$.
Alternatively, one may simply consider vector unparticles without
scalar unparticles.  These cases provide natural mechanisms to
suppress $c_2$, and would imply that the conformal window extends down
to very low energies.

\begin{figure}
\resizebox{3.65 in}{!}{\includegraphics{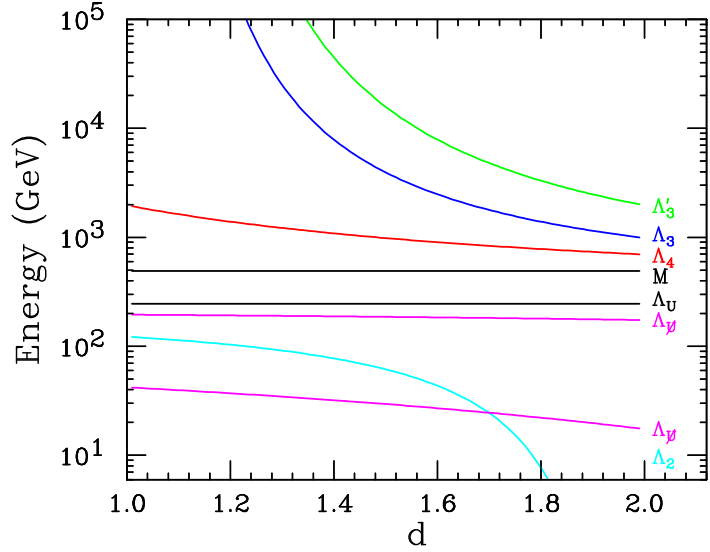}} 
\caption{Energy scales in the minimal unparticle model as functions of
$d$, assuming $\LambdaU = v \simeq 246~\gev$, $M=2v$, and $\duv =
3$. The two lines for $\LambdanotU$ are for $c_2=1$ (upper) and
$c_2=0.01$ (lower).
\label{fig:scales2}
}
\end{figure}

For whatever reason, either assuming fine-tuning or one of the more
natural possibilities noted above, one may consider $c_2 \ll 1$.  We
can then take $\Lambda_U \sim M$, implying $\Lambda_4 \sim \Lambda_3
\sim \Lambda'_3 \sim \Lambda_U$. This possibility is illustrated in
\figref{scales2}, where we have taken $M = 2\LambdaU = 2v$. For $c_2
\sim 0.01$, the conformal window may extend down to $\sim
10~\gev$. (See also \figref{scales10}.)  At the same time, $\Lambda_4,
\Lambda_3, \Lambda'_3 \sim \tev$, a scale at which colliders and other
high energy experiments can potentially probe fermion and gauge boson
couplings to unparticles.

\section{Scalar Unparticles}
\label{sec:scalar}

In this section, we assume that the conformal hidden sector couples
through a single scalar operator ($S = 0$). We consider $r\sim 1$ (and
therefore also $\Lambda_2\sim \Lambda_3\sim \Lambda_4\sim \LambdaU$),
and explore two classes of signatures in the conformal window at
energies near the weak scale $v$.  We show that scalar unparticles can
modify gauge couplings, possibly leading to an exotic signal.  Scalar
unparticles can also be seen in modifications to cross-sections at
high energy colliders.

\subsection{Contributions to Gauge Coupling Evolution}

Consider an unparticle operator $O$ that couples both to Higgs bosons
(as in \cite{Fox:2007sy}) and to gauge fields
\begin{equation}
\label{eq:Ohiggsgauge} 
c_2 \Lambda_2^{2-d} O H^2 + c_4^F \frac{O F^2}{\Lambda_4^d} \ .
\end{equation}
Below the scale of electroweak symmetry breaking, the first of these
interactions turns into a tadpole for $O$, which leads to the breaking
of the conformal invariance at the scale
$\LambdanotU$~\cite{Fox:2007sy}. Generically, $O$ obtains a vev of the
same order of magnitude, modifying the gauge kinetic term to
\begin{equation}
\left[\frac{1}{4}+ \mathcal{O}(1)
\left(\frac{\LambdanotU}{\Lambda_4}\right)^d \right] F^2\ ,
\end{equation}
where we have assumed that $c_4^F\sim 1$ and $\LambdanotU$ depends on
$c_2$.  Alternatively, one could obtain the same result by integrating
out $O$ at the threshold $\mu=\LambdanotU$.  This can be interpreted
as a threshold correction to the gauge coupling
\begin{equation}
\label{eq:couplingmod}
\left.\Delta \left(\alpha^{-1} \right) \right|^{\mu_1}_{\mu_2}\sim
4 \alpha^{-1} \left(\frac{\LambdanotU}{\Lambda_4}\right)^d, \quad 
\mu_1<\LambdanotU<\mu_2 \ .
\end{equation}
Thus it is possible to probe scales of the unparticle physics by
comparing values of the gauge coupling above and below
$\LambdanotU$. 

{}From the phenomenological perspective, one is most interested in the
case where the conformal invariance of the unparticle sector is only
broken below the electroweak scale, $\LambdanotU<M_Z$. Existing
measurements of the fine structure constant at zero energy and at the
$Z$ pole are consistent with the standard model renormalization group
evolution within experimental and theoretical uncertainties.  In this
comparison, the largest uncertainty arises from the value of the
coupling at the $Z$ pole~\cite{Yao:2006px}
\begin{equation}
\label{eq:alphamz}
\alpha^{-1} (M_Z) = 127.918\pm 0.018 \ .
\end{equation}
Comparing \eqsref{eq:couplingmod}{eq:alphamz} we find that the scales
of unparticle physics must be quite large.  For example, taking
$d=1.5$ we find
\begin{equation}
\left(\frac{\LambdanotU}{\Lambda_4}\right) \alt 10^{-3} \ . 
\end{equation}

Written in this form, constraints on the fundamental scales of the
unparticle sector implicitly depend on several parameters: $c_2, r,
s$, and $d_{UV}$.  Thus, it is useful to rephrase this result in terms
of the required fine-tuning of $c_2$. Choosing values of the remaining
parameters as in \figref{scales10} allows for $c_2$ of the order of a
one-loop contribution, but as a result,
$\Lambda_3^\prime>\Lambda_3>\Lambda_4\sim 25~\tev$.  If we choose the
same parameters as in \figref{scales2}, we find $c_2<7\cdot 10^{-7}$,
which is significantly smaller that its natural one loop value.  As a
final example, consider the situation where all scales in the
unparticle sector are comparable. In this case, $c_2<1.3 \cdot 10^{-5}
(\LambdaU/5~\tev)^2$ is required, again a significant fine-tuning.

\subsection{Enhancements of Collider Cross Sections}

Scalar unparticles may also be probed more directly by looking at
their effects on high energy processes. To analyze these effects, we
conservatively consider only gauge-invariant operators that are also
$B$-, $L$-, and flavor-conserving. In general, one could also consider
operators that violate one or more of these global symmetries ---
these will be much more stringently constrained.

For scalar unparticles, at leading order in $\Lambda$ there are
two types of interactions with standard model fermions:
\begin{equation}
\frac{e c_4^f}{\Lambda_4^d} O H \overline{f_L} f_R \ , \quad 
\frac{e c'_4{}^{f_{L,R}}}{\Lambda_4^d}
\partial^{\mu} O \overline{f_{L,R}} \gamma_{\mu} f_{L,R}
\ ,
\label{LOscalar}
\end{equation}
where we have inserted the electromagnetic coupling $e$ for future
convenience.  When electroweak symmetry is broken, the first term
generates $O \overline{f_L} f_R$ couplings proportional to $v$.  For
the second term, integrating by parts, the vector contribution
vanishes and the axial vector contribution is proportional to $m_f$.
Since $m_f \ll v \simeq 246~\gev$ for all but the top quark, we expect
the first class of operators to dominate and focus on those operators
here.

The scalar unparticle Feynman rules are
\begin{eqnarray}
O \overline{f_L} f_R \ \text{vertex:}
&& i e \frac{c^f_4 v}{\Lambda_4^d} P_R \ , \nolabel \\
O \ \text{propagator:} && \frac{i}{(q^2)^{2-d}} B_d \, ,
\label{Feynmanscalar}
\end{eqnarray}
where~\cite{Cheung:2007ue,Georgi:2007si}
\begin{equation}
B_d \equiv A_d \frac{\left(e^{-i\pi} \right)^{d-2}}{2 \sin d\pi} \
, \quad A_d \equiv \frac{16 \pi^{5/2} \Gamma(d+\frac{1}{2})}
{(2\pi)^{2d} \, \Gamma(d-1)\, \Gamma(2d)} \ . \label{bd}
\end{equation}

The $O$ interactions contribute to fermion pair production at
colliders through $f_1 \overline{f_1} \to O \to f_2 \overline{f_2}$.
In contrast to the case of vector unparticles discussed below, scalar
unparticles do not interfere with standard model $\gamma$ and $Z$
diagrams.  Specializing to the case of massless initial state
fermions, we find
\begin{equation}
\left. \frac{d\sigma^O}{dx} \right|_{\text{CM}} = N \frac{\pi
\alpha^2 s}{2} v_f \frac{1+v_f^2}{2} \frac{|B_d c_4^1 c_4^2|^2
v^4}{s^{4-2d} \Lambda_4^{4d}} \ , \label{scalarsigma}
\end{equation}
where $N$ is the numerical spin/color factor from averaging over
initial states and summing over final states, $\sqrt{s}$ is the
(parton-level) center-of-mass energy, $v_f = (1 - 4
m_{f_2}^2/s)^{1/2}$ is the final state particles' velocity in the
center-of-mass frame, and $x = \cos \theta$, where $\theta$ is the
angle between the incoming $f_1$ and outgoing $f_2$.  Scalar $O$
unparticles simply produce an isotropic increase in the cross section.

\begin{figure}
\resizebox{3.65 in}{!}{
\includegraphics{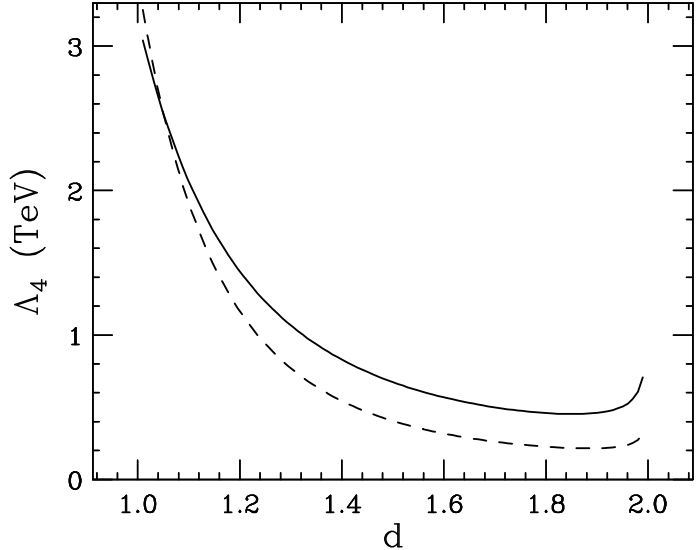}
} 
\caption{Lower bounds from LEP/SLC~\cite{Alcaraz:2006mx} (solid) and
JADE~\cite{Bartel:1985cs} (dashed) on the scalar $O$ interaction scale
$\Lambda_4$ from processes $e^+ e^- \to \mu^+ \mu^-$ as a function of
the dimension $d$ of the $O$ operator.
\label{fig:emuscalar}
}
\end{figure}

The enhancements of \eqref{scalarsigma} are constrained by many
experiments and many observables, including total cross sections and
forward-backward asymmetries.  For simplicity, we focus here on total
cross sections.  Experiments set upper bounds $\Delta
\sigma_{\text{exp}}$ on new physics contributions to $f_1
\overline{f_1} \to f_2 \overline{f_2}$ at center-of-mass energy
$\sqrt{s}_{\text{exp}}$.  This implies the bound
\begin{equation}
\Lambda_4 > \sqrt{s}_{\text{exp}} \left[ \frac{\pi}{16} v_f
\frac{1+v_f^2}{2} |B_d|^2 \frac{v^4}{ \sqrt{s}_{\text{exp}}^{\ 6}
\Delta \sigma_{\text{exp}}} \right]^{\frac{1}{4d}} \, ,
\end{equation}
where we have assumed the minimal set of non-zero couplings to see an
effect and taken $e^2 |c_4^1 c_4^2| = 2 \pi$ for consistency with the
compositeness literature, which we discuss in \secref{contact}.  Lower
bounds from a variety of processes and experiments are shown in
\figref{emuscalar} and \tableref{scalarbounds}.  Note that, since we
are considering small enhancements to standard model cross sections,
constraints from the $Z$ pole are not significant. The most stringent
bounds are from $e\mu$, and these vary from $\Lambda_4 > 2.1~\tev$ at
$d = 1.1$ to $\Lambda_4 > 460~\gev$ at $d=1.9$.  Of course, for a
given model, the cross section is enhanced at all energies; these
bounds could be improved by combining cross section and $A_{FB}$ data
from many different center-of-mass energies and experiments.

\begin{table}
\begin{tabular}{|c|c|c|c|c|c|c|}
 \hline \multirow{2}{*}{\ $f_1 f_2$ \ }
& \multirow{2}{*}{\ Experiment \ } & \multirow{2}{*}{\
$\sqrt{s}_{\text{exp}}~[\gev]$ \ } & \multirow{2}{*}{\
$\Delta \sigma_{\text{exp}}~[\fb]$ \ }
& \multicolumn{3}{c|}{\ Lower Bound on $\Lambda_4~[\gev]$ \ } \\
\cline{5-7} &&&& {\quad $d=1.1$ \quad } &
{\quad $d=1.5$ \quad } & {\quad $d=1.9$ \quad} \\
\hline \multirow{2}{*}{$e \mu$}
& LEP/SLC~\cite{Alcaraz:2006mx} & 189 & 76 & 2100 & 670 & 460 \\
\cline{2-7}
& JADE~\cite{Bartel:1985cs} & 34.6 & 1600 & 1900 & 400 & 220 \\
\hline \multirow{2}{*}{$e \tau$}
& LEP/SLC~\cite{Alcaraz:2006mx} & 189 & 100 & 1900 & 640 & 440 \\
\cline{2-7}
& JADE~\cite{Bartel:1985cs} & 34.6 & 2400 & 1700 & 380 & 200 \\
\hline \multirow{2}{*}{$e q$}
& LEP/SLC~\cite{Alcaraz:2006mx} & 189 & 240 & 1600 & 560 & 400 \\
\cline{2-7}
& TOPAZ~\cite{Miyabayashi:1994ej} & 57.8 & 4700 & 1200 & 340 & 210 \\
\hline \multirow{2}{*}{$e b$}
& LEP/SLC~\cite{Alcaraz:2006mx} & 189 & 140 & 1800 & 610 & 430 \\
\cline{2-7}
& VENUS~\cite{Abe:1993xr} & 58.0 & 3100 & 1400 & 360 & 220 \\
\hline
\end{tabular}
\caption{Lower bounds on $\Lambda_4$ from scalar $O$ interactions, for
4 pairs of fermion species $f_1 f_2$ and 3 representative values of
dimension $d$.  These are derived from $\Delta \sigma_{\text{exp}}$,
the upper bound on new physics contributions to $f_1 \overline{f_1}
\to f_2 \overline{f_2}$ at center-of-mass energy
$\sqrt{s}_{\text{exp}}$ at the experiments
named. \label{table:scalarbounds} }
\end{table}

\begin{figure}
\resizebox{3.65 in}{!}{\includegraphics{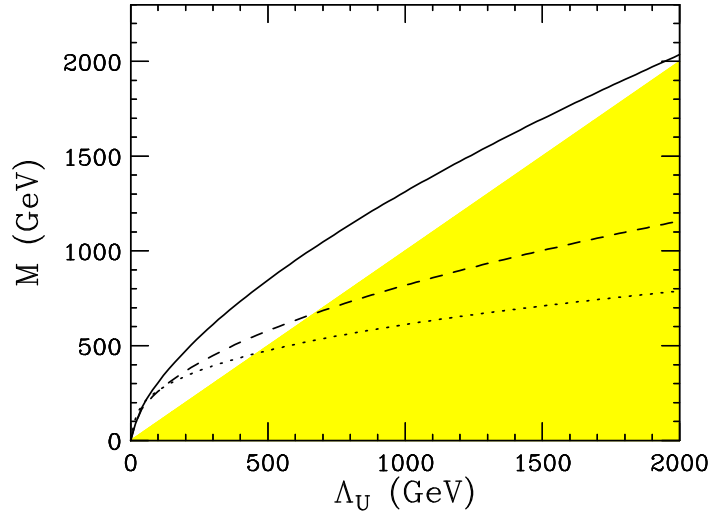}} 
\caption{Bounds from $e^+ e^- \to \mu^+ \mu^-$ on the fundamental
parameter space $(\LambdaU, M)$ for a scalar unparticle operator with
$\duv = 3$, and $d=1.1$ (solid), 1.5 (dashed), and 1.9 (dotted).  The
regions below the contours are excluded.  The shaded region is
excluded by the requirement $M > \LambdaU$.
\label{fig:lumscalar}
}
\end{figure}

The bounds we have derived may also be recast in terms of bounds on
the fundamental parameter space of the minimal models discussed in
\secref{models}.  In \figref{lumscalar} we consider minimal models
with $S=0$, $\duv=3$, and $c_2 \alt 0.01$, and show constraints from
$e^+e^- \to \mu^+ \mu^-$ in the $(\LambdaU, M)$ plane for $d=1.1, 1.5,
1.9$.  We see that the constraints have a significant $d$ dependence.
For $d=1.9$, the constraints primarily exclude parameter space that is
already excluded by the requirement $M > \LambdaU$.  However, for
$d=1.1$, the disfavored region is extended and excludes $M = \LambdaU$
up to 2 TeV.

\section{Vector Unparticles}

\subsection{Differential Cross Sections}

For vector unparticles, the leading coupling to fermions is through
the interactions
\begin{equation}
\frac{e c_3^{f_{L,R}}}{\Lambda_3^{d-1}} O^{\mu} 
\overline{f_{L,R}} \gamma_{\mu} f_{L,R} \ . \label{LO}
\end{equation}
These affect standard model fermion production through virtual
$O^{\mu}$ effects, and also through real $O^{\mu}$ production.  The
new Feynman rules involving $O^{\mu}$ particles are
\begin{eqnarray}
O^{\mu} \overline{f_i} f_i \ \text{vertex:} && i e
\frac{c_3^{f_i}}{\Lambda_3^{d-1}} P_i \gamma^{\mu} \ , \nolabel \\
O^{\mu} \ \text{propagator:} && \frac{i}{(q^2)^{2-d}} B_d \left(
- g_{\mu\nu} + \frac{q_{\mu} q_{\nu}}{q^2} \right) \ ,
\label{Feynmanvector}
\end{eqnarray}
where $B_d$ is as defined in \eqref{bd}, and we have assumed
$\partial_{\mu} O^{\mu} = 0$.

We focus here on new contributions to $f_i \overline{f_i} \to f_2
\overline{f_2}$ with $O^{\mu}$ unparticles in the $s$-channel, which
interfere with the corresponding photon and $Z$ diagrams.  Again
assuming massless initial state fermions, the resulting total
differential cross section is
\begin{eqnarray}
\left. \frac{d\sigma}{dx} \right|_{\text{CM}} &=& N \frac{\pi
\alpha^2 s}{2} \sum_{\stackrel{i,j=}{\gamma,Z,O}} \Delta_i
\Delta_j^* \left[ X^{ij}_1 X^{ij}_2 (1+v_f^2 x^2) + Y^{ij}_1
Y^{ij}_2 2 v_f x + X^{ij}_1 Z^{ij}_2 (1-v_f^2) \right] ,
\label{sigma}
\end{eqnarray}
where
\begin{eqnarray}
X^{ij}_k &=& Q_{k_L}^i Q_{k_L}^{j\,*} + Q_{k_R}^i Q_{k_R}^{j\,*} \\
Y^{ij}_k &=& Q_{k_L}^i Q_{k_L}^{j\,*} - Q_{k_R}^i Q_{k_R}^{j\,*} \\
Z^{ij}_k &=& Q_{k_R}^i Q_{k_L}^{j\,*} + Q_{k_L}^i Q_{k_R}^{j\,*} \ .
\end{eqnarray}
The vertex factors are
\begin{equation}
Q^Z_{f_i}
=\frac{I_{f_i} - Q^{\gamma}_{f_i}\sin^2 \theta_W} {\sin \theta_W
\cos \theta_W} \qquad\qquad 
Q^O_{f_i} =\frac{c_3^{f_i}}{\Lambda_3^{d-1}} \ , \\
\end{equation}
where $Q^{\gamma}_{f_i}$ and $I_{f_i}$ are the electric charge and
isospin of $f_i$.  The propagator factors are
\begin{equation}
\Delta_{\gamma} = \frac{1}{q^2} \qquad\qquad 
\Delta_Z = \frac{1}{q^2-m_Z^2 + i m_Z \Gamma_Z} \qquad\qquad 
\Delta_O = \frac{B_d}{(q^2)^{2-d}} \ . \label{props}
\end{equation}
The remaining quantities are as defined below \eqref{scalarsigma}.

\subsection{Bounds from Effective Contact Interactions}
\label{sec:contact}

As can be seen from the discussion above, for fixed $\sqrt{s}$, the
effects of vector-like $O^{\mu}$ particles are identical to (possibly
complex) shifts in $\gamma, Z$ couplings.  Bounds on, say, $Z$
couplings from interactions $f_1 \overline{f_1} \to f_2
\overline{f_2}$, are therefore bounds on the scale $\Lambda$ of
$O^{\mu}$ interactions.

A less precise, but more convenient, correspondence is between
$O^{\mu}$ interactions and contact interactions.  For data collected
at a fixed center-of-mass energy $\sqrt{s}_{\text{exp}}$, $O^{\mu}$
vertices induce an effective four-fermion contact interaction
\begin{equation}
\frac{e^2 c_3^1 c_3^2 B_d}{\Lambda_3^{2d-2} s_{\text{exp}}^{2-d}}
\overline{f_1} \gamma^{\mu} f_1 \overline{f_2} \gamma_{\mu} f_2 \ . 
\label{Ocontact}
\end{equation}
This can be compared to the operator
\begin{equation}
\frac{\eta g^2}{2 \Lambda_c^2} \overline{f_1} \gamma^{\mu} f_1
\overline{f_2} \gamma_{\mu} f_2 \ , \label{contact}
\end{equation}
which has been studied extensively in the context of quark and lepton
compositeness.  We may therefore derive bounds on $O^{\mu}$
interactions from bounds on the operators of \eqref{contact}.

The propagator factor $B_d$ is complex.  If the $c_3^{f_i}$
coefficients are assumed real, the phase in $B_d$ has interesting
consequences.  For example, for $d=1.5$, $B_d$ is imaginary, and so at
the Z pole, the $O^{\mu}$ diagram interferes fully with the $Z$
diagram, but not with the $\gamma$ diagram.  This and other
interesting consequences of the propagator phase have been discussed
previously in Ref.~\cite{Georgi:2007si}.

The coefficients $c_3^{f_i}$ may be complex, however, and so the
operators of \eqref{Ocontact} have an unknown phase.  This ambiguity
is not unique to unparticles --- the operators of \eqref{contact} also
have, in principle, complex coefficients.  In the compositeness
literature, this uncertainty is partially accounted for by deriving
bounds for $\eta = \pm 1$, thereby allowing either constructive or
destructive interference.  In our case, we will derive bounds assuming
$c_3^1 c_3^2 B_d$ is real and positive.  Bounds for other phases, or
incorporating the variation in phase with $d$, will differ slightly,
just as bounds on \eqref{contact} depend on the sign of $\eta$.

The resulting bound on the scale of $O^{\mu}$ interactions is
\begin{equation}
\Lambda_3 > |B_d|^{\frac{1}{2d-2}} \Lambda_{\text{exp}} \left(
\frac{\Lambda_{\text{exp}}}{\sqrt{s}_{\text{exp}}}
\right)^{\frac{2-d}{d-1}} \ , \label{bound}
\end{equation}
where $\Lambda_{\text{exp}}$ is the bound on the compositeness scale
$\Lambda_c$ in \eqref{contact} resulting from data taken at
center-of-mass energy $\sqrt{s}_{\text{exp}}$, and we have followed
the conventions of the fermion compositeness
literature~\cite{Eichten:1983hw,Yao:2006px} in assuming the minimal
set of non-zero couplings to see an effect and setting $e^2 |c_3^1
c_3^2 | = g^2/2 \ ( = 2 \pi)$.

Equation (\ref{bound}) has several interesting features.  First, given
$\Lambda_{\text{exp}} > \sqrt{s}_{\text{exp}}$, the bound becomes
increasingly stringent as $d \to 1$.  This is as it should be --- in
this limit, the operator of \eqref{LO} becomes almost renormalizable,
and so less sensitive to $\Lambda_3$.  Second, for equivalent
$\Lambda_{\text{exp}}$, the bound is more stringent for {\em lower}
$\sqrt{s}_{\text{exp}}$.  As a result, constraints from experiments
now far from the energy frontier, but still above $\LambdanotU$, the
scale of conformal symmetry breaking, are in some cases the leading
constraints.

\begin{figure}
\resizebox{3.65 in}{!}{
\includegraphics{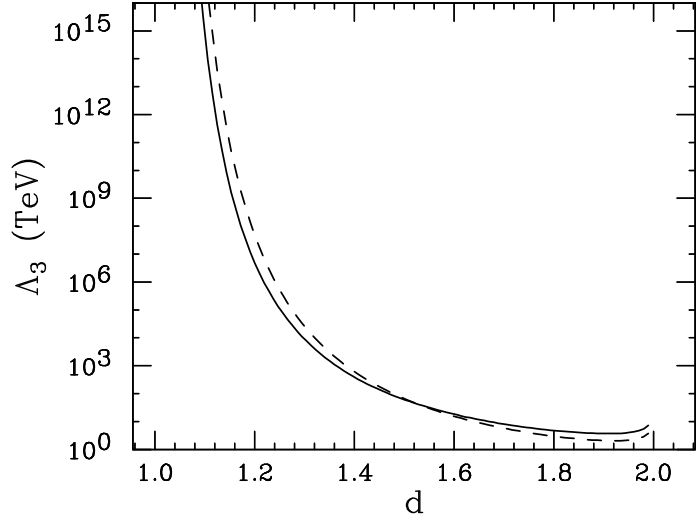}
} 
\caption{Lower bounds from L3~\cite{Acciarri:2000uh} (solid) and
JADE~\cite{Bartel:1985cs} (dashed) on the vector $O^{\mu}$ interaction
scale $\Lambda_3$ from processes $e^+ e^- \to \mu^+ \mu^-$ as a
function of the dimension $d$ of the $O^{\mu}$ operator.
\label{fig:emuvector}
}
\end{figure}

We present bounds on $\Lambda_3$ in \figref{emuvector} and
\tableref{vectorbounds}.  In every case, we conservatively choose
$\sqrt{s}_{\text{exp}}$ to be the maximum center-of-mass energy at
which the relevant data were taken.  This is a conservative
assumption, but the bounds are not very sensitive to it, especially
for $d$ near 2, where the constraints on $\Lambda_3$ are least
stringent. For example, at $d=1.9$, taking $\sqrt{s}_{\text{exp}} =
136~\gev$ instead of 189 GeV for LEP2 bounds strengthens the bound on
$\Lambda_3$ by only 4\%.

\begin{table}
\begin{tabular}{|c|c|c|c|c|c|c|}
 \hline \multirow{2}{*}{\ $f_1 f_2$ \ }
& \multirow{2}{*}{\ Experiment \ } & \multirow{2}{*}{\
$\sqrt{s}_{\text{exp}}~[\gev]$ \ } & \multirow{2}{*}{\
$\Lambda_{\text{exp}}~[\tev]$ \ }
& \multicolumn{3}{c|}{\ Lower Bound on $\Lambda_3~[\tev]$ \ } \\
\cline{5-7} &&&& {\quad $d=1.1$ \quad } &
{\quad $d=1.5$ \quad } & {\quad $d=1.9$ \quad} \\
\hline \multirow{2}{*}{$e_L \mu_L$}
& L3~\cite{Acciarri:2000uh} & 189 & 8.5 & $9.1 \times 10^{14}$ & 61 & 3.7 \\
\cline{2-7}
& JADE~\cite{Bartel:1985cs} & 46.8 & 4.4 & $3.6 \times 10^{17}$ & 66 & 2.1 \\
\hline \multirow{2}{*}{$e_L \tau_L$}
& L3~\cite{Acciarri:2000uh} & 189 & 5.4 & $9.7 \times 10^{12}$ & 25 & 2.2 \\
\cline{2-7}
& JADE~\cite{Bartel:1985cs} & 46.8 & 2.2 & $3.5 \times 10^{14}$ & 16 & 1.0 \\
\hline \multirow{2}{*}{$e_L q_L$}
& OPAL~\cite{Abbiendi:2003dh} & 207 & 8.2 & $2.8 \times 10^{14}$ & 52 & 3.5 \\
\cline{2-7}
& TOPAZ~\cite{Adachi:1991pq} & 57.9 & 1.2 & $1.2 \times 10^{11}$ & 4.0 & 0.48 \\
\hline \multirow{2}{*}{$e_L b_L$}
& ALEPH~\cite{Barate:1999qx} & 183 & 5.6 & $1.9 \times 10^{13}$ & 27 & 2.3 \\
\cline{2-7}
& CELLO~\cite{Behrend:1990gb} & 43 .0 & 1.1 & $7.3 \times 10^{11}$ & 4.5 & 0.45 \\
\hline
\end{tabular}
\caption{Lower bounds on $\Lambda_3$, the scale of vector $O^{\mu}$
interactions, for 4 pairs of fermion species $f_1 f_2$ and 3
representative values of dimension $d$.  These are derived from
$\Lambda_{\text{exp}}$, the lower bound on the scale of four-fermion
contact interactions derived from data with maximum center-of-mass
energy $\sqrt{s}_{\text{exp}}$ at the experiments named.
\label{table:vectorbounds} }
\end{table}

The results given in \figref{emuvector} and \tableref{vectorbounds}
illustrate the features noted above.  For low $d$, the bounds on
$\Lambda$ rise quickly, and are in some cases above the Planck scale
for $d=1.1$.  At the same time, even for $d$ near 2, the lower bound
on $\Lambda_3$ is at least 2 TeV in all channels considered.  We also
see that many of the leading bounds for low $d$ arise from data at
$\sqrt{s}_{\text{exp}} \sim 50~\gev$, far from LEP2 energies.

\begin{figure}
\resizebox{3.65 in}{!}{\includegraphics{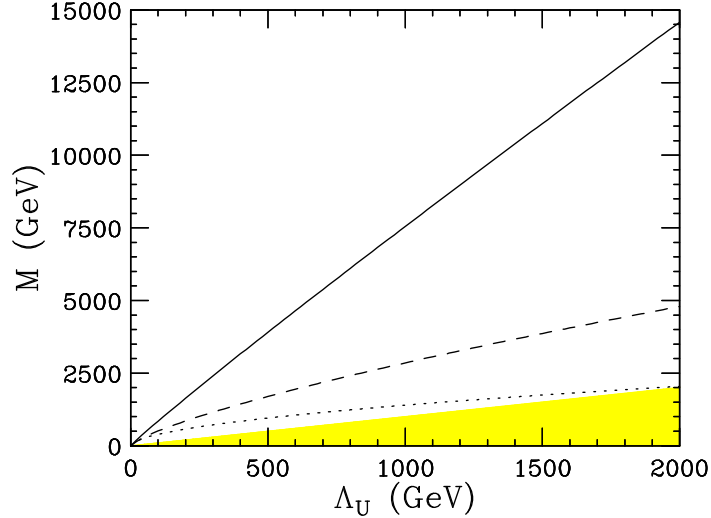}} 
\caption{Bounds from $e^+ e^- \to \mu^+ \mu^-$ on the fundamental
parameter space $(\LambdaU, M)$ for a vector unparticle operator with
$\duv = 3$, and $d=1.1$ (solid), 1.5 (dashed), and 1.9 (dotted).  The
regions below the contours are excluded.  The shaded region is
excluded by the requirement $M > \LambdaU$.
\label{fig:lumvector}
}
\end{figure}

As in \secref{scalar}, we also present the bounds of
\tableref{vectorbounds} as constraints in the fundamental parameter
space of minimal models in \figref{lumvector}.  Bounds in the
$(\LambdaU, M)$ plane are not singular as $d \to 1$.  The singularity
in $\Lambda_3$ is artificial --- the physically relevant quantity is
$\Lambda_3^{d-1}$ not $\Lambda_3$ --- and this is removed by
considering the fundamental parameters.  In contrast to the scalar
case, we find that the bounds are far stronger than the consistency
requirement $M > \LambdaU$.  Vector unparticle effects are enhanced by
interference with the standard model, in contrast to the scalar
unparticle case.
 
We have given only a sampling of possible bounds, corresponding to
compositeness bounds with $\eta = 1$ with $LL$ couplings only.  Bounds
for $\eta = -1$ and different chiralities are simple to derive.  These
results may also be extended to other $f_1 f_2$ pairs, although the
definition of $\sqrt{s}_{\text{exp}}$ is less well-defined for bounds
from hadron-hadron or lepton-hadron interactions and also for $e_{L,R}
e_{L,R}$, where $t$-channel effects are present.

\section{Conclusions}

Unparticles from conformal hidden sectors provide qualitatively new
signals for new physics.  We have considered the theoretical
framework of minimal models, both to include the constraint of low
conformal symmetry breaking and to clarify the relationship of the
various scales that enter phenomenological analyses.

We considered couplings of scalar unparticles to standard model
operators of dimension 2 and 4 and vector unparticles to dimension 3
operators. The mass scales $\Lambda_n$ of these couplings are related
to the onset of scale invariance $\LambdaU$ by specific powers of a
common factor $r\le 1$. Results are presented for two values $r=0.5$
and $r=0.1$. We find that the requirement of a significant conformal
window places strong constraints on models.

These considerations suggest that the most robust probes of unparticle
effects must come from high energies. With this in mind, we then
considered some of the most promising probes of unparticle effects.
We derived bounds on the scales for both scalar and vector unparticles
from precision $e^+ e^-$ data at center-of-mass energies $\sqrt{s}
\approx 30~\gev$ to 200 GeV. These bounds were determined for a number
of representative channels and presented both in terms of the
phenomenological parameters $\Lambda_n$ and in terms of constraints on
the fundamental mass parameters $M$ and $\LambdaU$.

The analysis of \secref{scales} implies that $\Lambda_4 < \Lambda_3$,
that is, that the characteristic mass scale for unparticle couplings
to standard model gauge bosons is lower than for unparticle couplings
to standard model fermions.  This suggests that stronger bounds than
the ones found here may be derived from gluon-gluon processes at
hadron colliders such as the Tevatron and LHC, or possibly from
enhancements to ultra-high energy cosmic ray and cosmic neutrino cross
sections.  These analyses are left to future work.

We likewise noted an exotic effect of scalar unparticles on gauge
coupling evolution. Application to the running of the fine structure
constant to $M_Z$ resulted in an unnaturally small value of the
coefficient of the coupling of dimension 2 operators, $H^2$, to scalar
unparticles. This disfavors scalar unparticles and suggests that only
vector ones couple to the standard model.

\section*{Acknowledgments}

The work of JLF is supported in part by NSF CAREER grant
No.~PHY--0239817, NASA Grant No.~NNG05GG44G, and the Alfred P.~Sloan
Foundation. AR is partially supported by NSF grant PHY--0354993.  JLF,
AR, and YS are partially supported by NSF grant PHY--0653656.




\begin{thebibliography}{99}

\bibitem{Georgi:2007ek}
  H.~Georgi,
  Phys.\ Rev.\ Lett.\  {\bf 98}, 221601 (2007)
  [arXiv:hep-ph/0703260].

\bibitem{Georgi:2007si}
  H.~Georgi,
  arXiv:0704.2457 [hep-ph].

\bibitem{Cheung:2007ue}
  K.~Cheung, W.~Y.~Keung and T.~C.~Yuan,
  arXiv:0704.2588 [hep-ph].

\bibitem{Luo:2007bq}
  M.~Luo and G.~Zhu,
  arXiv:0704.3532 [hep-ph].

\bibitem{Chen:2007vv}
  C.~H.~Chen and C.~Q.~Geng,
  arXiv:0705.0689 [hep-ph].

\bibitem{Ding:2007bm}
  G.~J.~Ding and M.~L.~Yan,
  arXiv:0705.0794 [hep-ph].

\bibitem{Liao:2007bx}
  Y.~Liao,
  arXiv:0705.0837 [hep-ph].

\bibitem{Aliev:2007qw}
  T.~M.~Aliev, A.~S.~Cornell and N.~Gaur,
  arXiv:0705.1326 [hep-ph].

\bibitem{Li:2007by}
  X.~Q.~Li and Z.~T.~Wei,
  arXiv:0705.1821 [hep-ph].

\bibitem{Duraisamy:2007aw}
  M.~Duraisamy,
  arXiv:0705.2622 [hep-ph].

\bibitem{Lu:2007mx}
  C.~D.~Lu, W.~Wang and Y.~M.~Wang,
  arXiv:0705.2909 [hep-ph].

\bibitem{Stephanov:2007ry}
  M.~A.~Stephanov,
  arXiv:0705.3049 [hep-ph].

\bibitem{Fox:2007sy}
  P.~J.~Fox, A.~Rajaraman and Y.~Shirman,
  arXiv:0705.3092 [hep-ph].

\bibitem{Greiner:2007hr}
  N.~Greiner,
  arXiv:0705.3518 [hep-ph].

\bibitem{Davoudiasl:2007jr}
  H.~Davoudiasl,
  arXiv:0705.3636 [hep-ph].

\bibitem{Choudhury:2007js}
  D.~Choudhury, D.~K.~Ghosh and Mamta,
  arXiv:0705.3637 [hep-ph].

\bibitem{Chen:2007qr}
  S.~L.~Chen and X.~G.~He,
  arXiv:0705.3946 [hep-ph].

\bibitem{Aliev:2007gr}
  T.~M.~Aliev, A.~S.~Cornell and N.~Gaur,
  arXiv:0705.4542 [hep-ph].

\bibitem{Mathews:2007hr}
  P.~Mathews and V.~Ravindran,
  arXiv:0705.4599 [hep-ph].

\bibitem{Zhou:2007zq}
  S.~Zhou,
  arXiv:0706.0302 [hep-ph].

\bibitem{Ding:2007zw}
  G.~J.~Ding and M.~L.~Yan,
  arXiv:0706.0325 [hep-ph].

\bibitem{Chen:2007je}
  C.~H.~Chen and C.~Q.~Geng,
  arXiv:0706.0850 [hep-ph].

\bibitem{Liao:2007ic}
  Y.~Liao and J.~Y.~Liu,
  arXiv:0706.1284 [hep-ph].

\bibitem{Strassler:2006im}
  M.~J.~Strassler and K.~M.~Zurek,
  arXiv:hep-ph/0604261;
  arXiv:hep-ph/0605193.

\bibitem{Patt:2006fw}
  B.~Patt and F.~Wilczek,
  arXiv:hep-ph/0605188.

\bibitem{Eichten:1983hw}
  E.~Eichten, K.~D.~Lane and M.~E.~Peskin,
  Phys.\ Rev.\ Lett.\  {\bf 50}, 811 (1983).

\bibitem{Yao:2006px}
  W.~M.~Yao {\it et al.}  [Particle Data Group],
  J.\ Phys.\ G {\bf 33}, 1 (2006).

\bibitem{Alcaraz:2006mx}
ALEPH, DELPHI, L3, OPAL, and SLD Collaborations, LEP Electroweak
Working Group, SLD Electroweak Group, and SLD Heavy Flavor Group,
  arXiv:hep-ex/0612034.

\bibitem{Bartel:1985cs}
  W.~Bartel {\it et al.}  [JADE Collaboration],
  Z.\ Phys.\  C {\bf 30}, 371 (1986).

\bibitem{Miyabayashi:1994ej}
  K.~Miyabayashi {\it et al.}  [TOPAZ Collaboration],
  Phys.\ Lett.\  B {\bf 347}, 171 (1995).

\bibitem{Abe:1993xr}
  K.~Abe {\it et al.}  [VENUS Collaboration],
  Phys.\ Lett.\  B {\bf 313}, 288 (1993).

\bibitem{Acciarri:2000uh}
  M.~Acciarri {\it et al.}  [L3 Collaboration],
  Phys.\ Lett.\  B {\bf 489} (2000) 81
  [arXiv:hep-ex/0005028].

\bibitem{Abbiendi:2003dh}
  G.~Abbiendi {\it et al.}  [OPAL Collaboration],
  Eur.\ Phys.\ J.\  C {\bf 33}, 173 (2004)
  [arXiv:hep-ex/0309053].

\bibitem{Adachi:1991pq}
  I.~Adachi {\it et al.}  [TOPAZ Collaboration],
  Phys.\ Lett.\  B {\bf 255}, 613 (1991).

\bibitem{Barate:1999qx}
  R.~Barate {\it et al.}  [ALEPH Collaboration],
  Eur.\ Phys.\ J.\  C {\bf 12}, 183 (2000)
  [arXiv:hep-ex/9904011].

\bibitem{Behrend:1990gb}
  H.~J.~Behrend {\it et al.}  [CELLO Collaboration],
  Z.\ Phys.\  C {\bf 51}, 149 (1991).

\end{thebibliography}
\end{document}